\documentstyle[sprocl]{article}

\def\be{\begin{equation}}
\def\ee{\end{equation}}
\def\bea{\begin{eqnarray}}
\def\eea{\end{eqnarray}}

\begin{document}
\title{SONIC-POINT MODEL FOR HIGH-FREQUENCY QPOs IN NEUTRON STAR
LOW-MASS X-RAY BINARIES}
\author{ M. COLEMAN MILLER }
\address{University of Chicago, Department of Astronomy and
Astrophysics,\\ 5640 S. Ellis Ave., Chicago, IL 60637, USA}
\author{ FREDERICK K. LAMB AND DIMITRIOS PSALTIS }
\address{University of Illinois, Department of Physics and\\
Department of Astronomy, University of Illinois at Urbana-Champaign\\
1110 W. Green St., Urbana, IL 61801-3080, USA}


\maketitle\abstracts{
Quasi-periodic brightness oscillations (QPOs) with frequencies in the
range 300--1200~Hz have been detected from at least nine
neutron star low-mass X-ray binaries. Here we summarize the
sonic-point model for these brightness oscillations, which we present in
detail elsewhere. If the sonic-point
interpretation of kilohertz QPOs is confirmed, measurements
of kilohertz QPO frequencies in low-mass X-ray binaries will
provide new bounds on the masses and radii of neutron stars
in these systems and new constraints on the equation of state
of matter at high densities.}

\section{Introduction}
Observations of neutron stars in low-mass X-ray binaries
using the Rossi X-ray Timing Explorer have revealed
that at least nine produce
quasi-periodic X-ray brightness oscillations (QPOs) with
frequencies $\nu_{\rm QPO}$ ranging from $\sim 300$~Hz to
$\sim 1200$~Hz (see, e.g., van der Klis et al. 1996; 
Strohmayer et al. 1996; Berger et al. 1996; Zhang et al.
1996). These high-frequency QPOs are remarkably coherent
($\nu_{\rm QPO}/\Delta\nu_{\rm QPO}$ up to $\sim$200) and
strong (rms amplitudes up to 20\%). In several sources
the frequencies are strongly positively correlated
with countrate. In
four sources, a single highly coherent QPO peak is seen 
during type I X-ray bursts. High-frequency QPOs have been
observed to occur in pairs in six sources.

In the sonic-point model of these high-frequency QPOs, 
(1)~the frequency of the higher frequency QPO in a pair
is the Keplerian
frequency at the point in the disk flow where the inward
radial velocity increases steeply with decreasing radius
and typically becomes supersonic, and
(2)~the frequency of the lower frequency QPO peak is the
beat frequency between the sonic-point Keplerian frequency
and the stellar spin frequency or its overtones. 
We emphasize that the sonic-point
model is {\it not} a magnetospheric beat-frequency model
(see Miller, Lamb, \& Psaltis 1996 [hereafter MLP] for
further discussion of the differences between the sonic-point
model and magnetospheric models). The sonic-point model is
based on earlier work on the effect of radiation forces on
disk accretion by neutron stars (Miller \& Lamb 1993, 1996).

\section{Overview of the Sonic-Point Model}
The sonic-point model builds on the 
standard picture of the Z and atoll sources,
in which the neutron star accretes gas from
a low-mass companion via a geometrically thin disk and
stresses within the disk create only a slow, subsonic inward
radial drift. In the sonic-point model, near the neutron star there is
a region of the disk flow in which the inward radial velocity
increases rapidly with decreasing radius, primarily because
(1)~radiation from near the star removes angular momentum
from the gas in the disk, allowing the gas to accelerate
radially, or---if radiation forces are sufficiently
weak---(2)~the gas drifts inside the marginally stable orbit
$R_{\rm ms}$. Because there is a sharp transition to supersonic
radial inflow in a very small radial distance, we refer to
this radius as the ``sonic point" radius $R_{\rm sonic}$.
We expect that at least some gas in the disk will penetrate
in to $R_{\rm sonic}$ provided that the magnetic field of the
neutron star is less than $\sim 10^{10}$~G, as in the Z
and atoll sources.

We expect that the disk flow will have local density inhomogeneities,
or ``clumps". Because the radial velocity outside
$R_{\rm sonic}$ is small, the inflow time from this region 
to the stellar surface
is large and hence clumps produced outside $R_{\rm sonic}$ are sheared
or dissipated before gas from them reaches the surface. 
In contrast, the inflow
time for clumps from near $R_{\rm sonic}$ is small, and hence clumps
formed in this region
are not completely sheared or dissipated before gas from them
impacts the
stellar surface. The impact of gas with the stellar surface
produces bright, arc-shaped footprints that
move around the star.  As seen at infinity, the frequency 
at which the pattern of arcs moves
around the star is the Keplerian frequency
$\nu_{\rm Ks}$ at the sonic point (see MLP). 
Hence, from the standpoint of a distant observer
with a line of sight inclined with respect to the disk axis,
the bright arcs are periodically occulted by the star at the
frequency $\nu_{\rm Ks}$, and the observer sees
brightness oscillations at a frequency $\nu_{\rm Ks}$. 
If the field is weak but not negligible, some of
the accreting gas may be channeled by the magnetic field onto
hot spots on the surface that rotate with the star. Because the
inflow of gas from the clumps is caused by radiation drag, the enhanced
radiation intensity from the hot spots modulates the mass accretion
rate at the beat frequency
between the sonic-point Keplerian frequency and the stellar
spin frequency or its overtones. In this case, a distant
observer will also see a modulation of the total luminosity at the beat
frequency.

As we explain in detail in MLP, the sonic-point model explains
naturally the high frequencies,
amplitudes, and coherences of kilohertz QPOs and their changes
with countrate, as well as
the numerous other observed correlations. The sonic-point model
is also consistent with the physical picture constructed
previously from observations of the spectra and low-frequency
variability of the Z and atoll sources.

\section{Predictions and Implications}

The sonic-point model makes a number of testable predictions, which
are discussed in detail in MLP. In particular, we do
not expect kilohertz QPOs of the type discussed here in black hole sources,
because in the sonic-point model the presence of a stellar surface is
essential. We expect that QPOs of this type will be very
weak or undetectable by current instruments from sources which show
strong periodic oscillations at their stellar spin frequencies
(pulsars), since these sources have strong magnetic fields and
hence at most a very 
a small fraction of the accreting gas will reach the stellar
surface without being forced by the magnetic field to corotate
with the neutron star.

If the sonic-point model is correct, 
observation of a high-frequency QPO provides 
upper limits to the mass and radius of the neutron star that are 
independent of the equation of state and, for an assumed equation
of state, yields estimates of the mass and radius of the source.
These constraints can in principle rule out equations of state,
particularly if high ($>$1300~Hz) QPO frequencies are observed,
thereby constraining the properties of dense matter.
\smallskip

This work was supported in part by
NSF grant AST 93-15133 and NASA grant 5-2925 at the University
of Illinois, NASA grant NAG 5-2868 at the University of Chicago,
and through the {\it Compton Gamma-Ray Observatory} Fellowship
Program, by NASA grant NAG 5-2687.


\section*{References}
\begin{thebibliography}{99} 
\bibitem{b96}Berger, M., et al., {\it Astrophys. J.},
469, L13 (1996).
\bibitem{ml93}Miller, M. C., \& Lamb, F. K.,
{\it Astrophys. J.}, 413, L43 (1993).
\bibitem{ml96}Miller, M. C., \& Lamb, F. K.,
{\it Astrophys. J.}, 470, 1033 (1996).
\bibitem{mlp}Miller, M. C., Lamb, F. K., \& Psaltis, D.,
{\it Astrophys. J.}, submitted (1996).
\bibitem{s96}Strohmayer, T. et al., {\it Astrophys. J.},
469, L9 (1996).
\bibitem{v96}van der Klis, M. et al., {\it Astrophys. J.},
469, L1 (1996).
\bibitem{z96}Zhang, W., Lapidus, I., White, N. E., \&
Titarchuk, L., {\it Astrophys. J.}, 469, L17 (1996).
\end{thebibliography}
\end{document}